# Formation and transformation of low density, onion-like carbon cages

*Shoaib Ahmad[1,2,*], Uzair Ahmed[2], Farrukh E. Mubarik[2], Tousif Hussain[2]*

[1]National Center for Physics (NCP), Shahdara Valley, Islamabad, Pakistan

[2] CASP, Government College University, Church Road, Lahore 54000, Pakistan

[*]E-mail address: sahmad.ncp@gmail.com (S. Ahmad)

## Abstract

A novel growth technique for low density, non-icosahedral carbon onion-like structures on Cu surface is described. The technique differs with the formation of carbon onions inside the $C_1^+$ implanted metal surfaces at high temperatures with densities ~ 2 g cm$^{-3}$ in the form of multishelled icosahedral fullerenes wrapped around $C_{60}$. We report spheroidal cage formation of shell with variable shell thickness, radii and curvature on $C_1^+$-irradiated Cu surfaces by the accumulating C atoms which emerge out of the edges of the implanted sheets and grids. Atom-by-atom C accreting structures grow along the edges, corners and crevices. These, in turn, form the open and closed multiple-shelled cages. Variable curvature is the most common feature of the growing structures. The growth at room temperature does not ensure sphericity of the cages due to the absence of high temperature annealing. These cages have low density (~10$^{-2}$-10$^{-3}$g cm$^{-3}$), hollow interiors and often exhibit the conspicuous cage-inside-cage structures. Another feature of the as-grown, non-spherical onions is the existence of spherical protrusions with different local radii within the same cage. Formation of cages with sizes ~ few nm-hundreds of nm depends upon the irradiation rate $C_1^+$ µm$^{-2}$ s$^{-1}$. In situ formation and transformations of the cages observed under 120 keV electron beam irradiation is described by using a nanoelastic model.

# 1. Introduction

In the experiments reported here we show that the energetic carbon ions ($C_1^+$) irradiated copper (Cu) grids and sheets with vertical holes let the carbon (C) atoms to emerge out and form graphitic structures along the edges. These structures include nano cages and stacks of graphitic layers. This technique is different as compared with C onion growing methods which utilized $C_1^+$ ion implantation of Cu at elevated temperatures. We do not utilize the implanted C ions to assemble within the host Cu matrix at high temperatures as was reported by other researchers [1-3]. A certain fraction of the implanted C atoms are believed to emerge, at the end of their implantation range, out of the irradiated Cu grids and sheets' edges and produce the graphitic structures outside the Cu matrix. We show in this communication the production of a variety of nano and micro-structures on the surface, at room temperature by utilizing smaller doses of C atoms. This is the fundamental difference in our growth technique with those of the other researchers where intense electron beams [4-8], arc discharge [9-12], laser ablation of graphite [13,14], high temperature combustion [15,16] and high pressure ball milling [17,18] methods were employed. Invariably, in all of these methods the higher number densities of the C atoms and the high thermal energy available to the C atoms ensure that the carbon onions produced are spherical, icosahedron-inside-icosahedron configurations i.e., $C_{60}@C_{240}@C_{540}@C_{960}@C_{1500}$... These C onions are without hollow interiors and have density approaching that of graphite (2.26 g cm$^{-3}$) [1-7]. Their reported diameters are in the range ~ few to many tens of nm. Various C onion producing techniques and the methods have been discussed in the context of the formation mechanisms [19-22]. The low solubility of C in Cu at all temperatures and the associated immiscibility gap is the basic reason for the choice of Cu for the growth of C nanostructures on the surface or within the solid [23-25]. In almost all ion implantation experiments, $C_1^+$ dose at the desired energy is the main parameter along with the high substrate temperature ~700-1000 K. The integrated dose of $C_1^+$ ~$10^{17}$ cm$^{-2}$ was reported for the formation of C onions [1-3, 24,25]. On the other hand, we have implanted ~$10^{15}$ $C_1^+$ cm$^{-2}$ at room temperature and observed that the implantation rate ($C_1^+$ cm$^{-2}$ s$^{-1}$) rather than the dose is the crucial parameter. The emerging C atoms at the end of their implantation range with almost no kinetic energy are the ingredients for the sp$^2$ bonds to form. This novel characteristic feature of energetic $C_1^+$ irradiated Cu was exploited for the present technique. Carbon onions with various shapes and sizes have been seen emerging along with the graphitic layers at all stages of C accretion and the subsequent nano and micro structure formation. Discontinuous or variable $C_1^+$ irradiation rate is likely to produce hollow cages that may be connected with each other or take the form that we identify as hollow C onions. Some of these hollow C onions have been observed to contain smaller ones: the configuration that we describe as the onion-inside-onion. During multiple irradiation sequence at successively increasing $C_1^+$ energies the initial batch of graphitic layers, sheets and shells are formed along the edges. These are lifted outwards during the next round of irradiation resulting in the growth underneath the earlier structures. All subsequent growth occurs beneath the earlier ones. This is another essential characteristic of the technique. A broad range of C nanostructures are observed. These include stacks of graphitic layers and cages formed along the 30 μm wide Cu grids used in transmission electron microscope (TEM). Similar structures emerged along the inner rings of 120 μm thick Cu sheets. The Cu grids and sheets were irradiated with 0.2-0.8 MeV $C_1^+$ ions. We present and discuss the results obtained using the scanning and transmission electron microscopes of the shapes and forms of these structures. Confocal microscopy was employed to confirm the graphitic nature of the objects and to study the mechanisms by which the C atoms emerge out of the Cu grids and sheets produce the graphitic structures. The structures formed with this technique are of very low densities, therefore, the detection by conventional tools like the energy dispersive spectroscopy, proton induced X-ray emission, Fourier transform infra red spectroscopy and X-ray diffraction could not identify the C cages or the graphitic layers that were clearly visible in the scanning and transmission electron microscopic micrographs. The Confocal images clearly identify the C structures on and along the sharp surfaces and edges of the irradiated Cu grids and sheets. Diffraction spots and rings by electrons in TEM further confirm the graphitic nature of the layers and cages.

## 2. Experimental

*2.1 Method*

The 2 MV Pelletron at GCU, Lahore was used for the implantation of $C_1^+$ at various energies and with the desired irradiation rates. The first set of targets was the 3 mm OD, 10 μm thick Cu grids with square and circular holes. These commercial grids are normally used to support nano particles for observation in TEM. Cu grids with 90 μm circular holes and 100 μm square holes were used for $C_1^+$ implantation. The second set of targets was punched out of the 120 μm thick Cu sheets with 3 mm OD and 1 mm ID circular holes. The edges of grids and the sheet holes were observed to be the regions where the implanted C atoms emerged to form the graphitic structures. Experiments were performed at energies between 0.2-0.8 MeV. The first series of experiments was done where the total dose was maintained at ~5(±1) x$10^{14}$ $C_1^+$ $cm^{-2}$. Detailed scanning and transmission electron microscope examination revealed that the irradiation rate ($C_1^+$ $cm^{-2}$ $s^{-1}$) is the fundamental parameter and must be kept constant during the irradiations at a given energy. It also helps to determine the densities of the structures so formed. During the second series of $C_1^+$ implantations, a step-wise increasing and decreasing energy irradiations in a staircase manner was employed. The same dose (~5 x$10^{14}$ $C_1^+$ $cm^{-2}$) at 0.2, 0.3, 0.4, 0.6 and 0.8 MeV was implanted with the cumulative dose of 6 x $10^{15}$ $C_1^+$ $cm^{-2}$. The initial batch of the targets irradiated with the single irradiation dose at the chosen energy provided us the essential information about the growth technique.

*2.2 Procedure*

The $C_1^+$ irradiation of Cu grids is performed on a composite target. The target consists of the grids placed on a Cu sheet which is held under another Cu sheet with 3 mm diameter holes to hold the grids in place and provide the mechanical support during irradiation. The irradiated grids and sheets were investigated with scanning electron and Confocal microscopes. The graphitic structures emerging out of the irradiated edges could be seen and confirmed in SEM and the Confocal microscopes. The second sequence of the experiments also utilized a set of three Cu sheets laid on top of each other. The 120 μm thick Cu sheets were cut as 15 mm x 45 mm strips. Two sets of seven holes, in the form of a hexagon around a central one, are punched in the two 15 mm x 15 mm sets. The 45 mm long sheets were folded in three overlapping layers; the top one had seven 3 mm diameter holes with a central and six peripheral ones, while the layer beneath had seven 1 mm diameter holes at approximately the same positions at the centers of the upper sheet's holes. The last sheet acted as the base plate and the recipient of the $C_1^+$ ions irradiating through the 1 mm diameter holes of the second set of Cu sheets. Each set of the three overlapping Cu sheets was used as the target to be irradiated with $C_1^+$ at a given energy. The combined sets were examined in the Confocal and the scanning electron microscopes before the 3 mm OD and 1 mm ID discs are punched out from the second Cu sheets for examination in TEM. Jeol's JSM-6480LV scanning electron microscope (SEM) was used to view and characterize the C structures that are formed as a result of $C_1^+$ irradiations. The implantation ranges of C ions in Cu with energies 0.2-0.8 MeV are ~0.2-0.7 μm which are much smaller than the thickness of the TEM's Cu grids. The graphitic objects are formed along the edges of the 10 μm thick grids and all along the inner edges of the 1 mm inner holes of sheets. The density of these objects, as will be discussed in the discussion section is very small. At the irradiation rate of ~$10^3$ $C_1^+$ $s^{-1}$ $\mu m^{-2}$ the structure formed in ~1000 s on Cu grid's bars of 30 μm x 30 μm catchment areas acquire a maximum ~$10^9$ C atoms. Even if all the irradiated C ions were to emerge as C atoms, at the end of their respective ranges in Cu (this being an extreme assumption), the density of objects observed is two to three orders of magnitude lower than the density of graphite. The details will be discussed later. Two sets of such C structures are shown in Fig. 1(a) and (b). Fig. 1(a) has the irradiated Cu grid with circular holes. One such hole is shown that has various objects ~10 μm size emerging out of the inner side of the 90 μm circular hole. Fig. 1(b) shows a 15-20 μm thick structure with about 600 μm length emerging out of the inner side of the 3 mm diameter hole's irradiated edge. The two insets in both the micrographs show the Confocal microscopic images of the grid and the hole in the sheet. The insets' images were obtained with Olympus Fluoview FV1000 microscope with λ=488 nm to highlight the C structures. The bright objects in the insets are the C microstructures against the dark Cu background.

Confocal microscopy with the appropriate laser light for identifying the C objects not only ensures the graphitic origin of the newly formed nano and micro-structures, it also helps in identifying the means by which these structures acquire C atoms. In Fig. 2(a), a quarter section of an irradiated 3 mm diameter square holed Cu grid's Confocal image is shown laced with the bright C objects. Similarly, Fig. 2(b) shows the central part of a 3 mm diameter Cu grid with circular holes. The circular holes are arranged in a simple cubic geometry with large areas of Cu in between that were also irradiated with $C_1^+$. Closer examination of these regions shows C structures of μm dimensions, in addition to the ones that are formed on and around the circular holes. Fig. 2(a) and (b) provide the route of formation of the graphitic structures by the C atoms emerging along the scratches and edges. Fig. 2(c) has the C decorated inner and outer edges of the 3 mm diameter hole in 120 μm thick Cu sheet. A bright region of ~20-30 μm running all along the periphery of the edge seems to be the irradiated carbon's repository. The inset in Fig. (2c) shows the Confocal image of a portion of the same edge where small objects can be seen protruding outwards and the differences of the Confocal contrasts reveal the presence of the surface in-homogeneities that have been differently decorated by the emerging C atoms.

## 3. The diversity of the structures produced by the emerging C atoms

*3.1 The gross features of the graphitic layers and cages*

Figure 3(a) is the scanning electron micrograph of a 1 mm diameter hole punctured in a 3 mm outer diameter Cu sheet to be investigated in the 120 kV TEM. The sheet was irradiated with an increasing energy sequence of $C_1^+$ at 0.2, 0.4, 0.6, 0.8 MeV followed by a decreasing energy down to 0.2 MeV. All irradiations were performed at the constant dose of $5 \times 10^{14}$ $C_1^+$ $cm^{-2}$. The cumulative dose of $C_1^+$ at all irradiating energies was $6 \times 10^{15}$ $C_1^+$ $cm^{-2}$. Irradiations were performed in such a way that the C structures could be formed in distinct stages and be distinguishable as separate batches. Our earlier investigations had shown that one could separate the C structures formed during irradiations at different energies. The formative process is dependent on the C atoms that emerge from inside the irradiated Cu sheets' edges and corners. All later additions of C atoms to the already formed layers or cages are from below. The graphitic structures formed in the earlier phases are pushed outwards by the later growth. This fact is demonstrated in Fig. 3(a) where the edges of the sheet are decorated with the outward emerging structures. One such formation is encircled on the right side of the circular hole decorated with C structures. One can see a large number of similar graphitic structures on the Cu base that acts as the beam stop. Figure 3(b) shows the TEM micrograph with enlarged structure shown encircled in 3(a). It has a rich variety of structures that are investigated and presented in this communication. The large object with about 10 μm x 15 μm dimensions is emerging outwards shown at the lower end. Its upper section as seen in SEM had shifted under the 120 keV electron beam to its present position. Its vertical posture can be seen in the circled section in the SEM micrograph 3(a). This object has overlapping, variable density regions. Most of the work presented in this communication is based on the two boxed sections labeled 1 and 2 in 3(b). These regions of ~ 1 $μm^2$ dimensions have an extensive range of hollow graphitic structures, open and closed are presented and discussed in the later setion.

*3.2 Structure and crystallinity of the graphitic cages*

Fig. 4 has the TEM micrographs of the two regions shown as 4(a) and 4(c). An outward edge near the boxed region 1 already shown in Fig. 3(b) is explored in 4(a). It occupies ~0.25 $μm^2$ area. Its diffraction pattern is shown in Fig. 4(b) at the black and white circled spot. The concentric rings are due to the irregular, randomly oriented graphitic layers. The rings are interspersed with few bright diffraction spots. A larger region of area ~ 2 $μm^2$ in 4(c) is from the centre of the box 1 of Fig. 3(b). It has a large range of irregular shaped structure with inter-connected layers that are studded with variable sized objects. The diffraction pattern from the black and white marked circles in 4(c) is shown in 4(d) as the bright rings laced with the diffraction spots indicative of the graphitic structures.

The variety of the cages present in 4(c) is further explored in Fig. 5. Fig. 5(a) has a larger view of the region with two boxed sections. Fig. 5(b) shows the upper left-hand white boxed structure (in 5(a)) enlarged to highlight the quasi-spherical objects studded on an underlying layered structure. These are identified as carbon onions of ~5-30 nm diameters. The central box in Fig. 5(a) is enlarged in 5(c). It focuses on a much larger collection of cages similar to the ones seen in Fig. 5(b). It has a collection of ~ a hundred C onions of various sizes, shapes and orientations. These seem to be connected with the underlying graphitic layers. The diffraction pattern from the center (Fig. 5(d)) shows couple of rings and many widely dispersed diffraction spots – the indications of the irregular shaped graphitic cages. The range of the clearly visible onions is 20-40 nm in 5(b) and 5(c) where a larger number of smaller ones are also present. The connectivity between the distinctly shaped and the separately positioned onions is presumed to be through the broken and the perforated layers. The diffraction spots from the centre of 5(c) are shown in the inset 5(d) that confirms the crystallinity of the graphitic shells; however, due to the in homogeneities of the C onion shapes and the associated variability of the curvature, the ordering of the spots is difficult. The range and extent of the shape and size irregularities will be further illustrated in the next section where the structural details of these C onions are discussed.

*3.3 Approximation of the density of a typical stack of graphitic layers*

Let us investigate the nature and the density of a typical graphitic stack of layers formed by the C atoms emergent from the irradiated Cu grids and sheets. Fig. 6(a) has such a stack whose diffraction pattern was shown in Fig. 4(b) comprising of rings. The stack with dimensions 600 nm x 300 nm x 300 nm has volume ~5 x$10^7$ nm$^3$. The irradiation rate was between 2-5 x $10^3$ $C_1^+$ μm$^{-2}$ s$^{-1}$ to deliver 6x$10^6$ $C_1^+$ μm$^{-2}$. Thus the stack shown in Fig. 6(a) of surface area of ~2 μm$^2$ with an assumed ten times larger catchment area will yield a maximum of about 2x$10^7$ C atoms during 1000 s irradiation. The maximum calculated density of the stack is about 2x$10^8$ C μm$^{-3}$. This is three orders of magnitude less than the density of graphite @2.26 g cm$^{-3}$ or 1.13x$10^{11}$ C μm$^{-3}$. Fig. 6(b) enlarges the boxed region of Fig. 6(a) with two outward protruding hollow shelled structures emerging from the underlying stack's outer periphery. The calculation of the approximate density of the stack can be used to evaluate the nature of these two half spherical cages superposed on a low density graphitic stack. It is emphasized again that these two shells were formed in the earliest phase of the stack's formation by the emerging C atoms, therefore, the structural similarities between these and the underlying stack can be assumed. The shells are ~3-5 nm thick with hollow interior. This seems to be the typical thickness of the overlapping layers in the stack.

# 4. The as-grown carbon onions

The nm sized objects present in figures 4, 5 and 6 were labeled as C onions. This is despite the fact that the densities of these cages formed during C ion implantation are much smaller than the density of graphite. In addition, our samples are kept at room temperature and the high temperature annealing required for the increased mobility, bond making and re-making is not available to the as-grown nanostructures formed by the accretion of the emerging C atoms out of the irradiated Cu sheets. In this section we investigate the detailed nature and the densities of the as-grown C onions and their manipulation with 120 keV electrons of TEM.

*4.1 Closed top, open bottom C onions*

Evidence is presented of C onions growing as a bunch of cages emerging out of the graphitic layered network. These C onions can be seen sticking to each other having partially open cages. Fig. 7(a) shows more than three dozen cages stuck together and emerging out of the graphitic layered base. The average cage size is ~30±5 nm. The ones that are hanging from the sides seem to be open in the section below through which C accretion may have taken place but the cages lifted outwards without the cage closure. Majority of the cages are in a state of imperfect closure. The upper part of the shells, on the other hand, seems more spherical and comparatively closed. Fig. 7(b) and its inset show the enlarged sections of the top and the bottom of the partially open cages. These cages seem to have been formed by accreting C from below resulting in the shell-inside-shell structures. The comparatively perfect hemispherical shells are due to the earlier stages of growth.

To investigate the causes of failure of the cage closure we performed an extended, intense electron beam irradiation experiment. Fig. 8 shows the effect of the continued irradiation with beam of intense 120 keV TEM electrons on the assembly of the interconnected, partially open C onions shown in Fig. 7. Six TEM micrographs are shown in Fig. 8(a-f). These were obtained during continuous electron beam irradiation. The C atoms were continuously being removed from the cages as the primary knock-on in head-on collisions with 120 keV electrons. The movement of the hind leg-like portion is counterbalanced by the rise of the front section with the eventual collapse of the whole bunch of cages towards the graphitic base to which it remains connected. The structures have shown maneuverability in terms of the reduction of the intra- and inter-cage bonds during 120 keV electron beam irradiation for 60 min. The removal of the C atoms in head-on collisions with 120 keV electrons leads to the creation of large number of vacancies in the cages' existing open structures. The collapse of the whole collection of cages shown in Fig. 8(f) indicates the breaking of significant number of bonds both within and across the interconnected C onions.

*4.2 The irregular shaped, close-caged, hollow C onions*

Fig. 9 shows the enlarged central portion of the graphitic layer studded with lot of cages that were already shown in Fig. 5(c). Six TEM topographs of ~ few dozen distinctly formed C onions are shown lying or connected to a graphitic base. Majority of these contain cages within cages that are described as the C onion-inside-onion. The successive micrographs show the structural changes of the onions under continuous 120 keV electron beam over 1200 s. A total of 44 micrographs were obtained. Six of these are shown here. The diameter of the white central ring around the scale bar on the fluorescent screen is 265 nm. The white circle in the centre of the topographs is to guide the eye in monitoring the irradiation induced changes. Here again one observes the changes induced by the energetic electrons knocking off C atoms leading to the shrinking and modifications of the size and shape of the cages. Sharp corners are smoothed; few of the cages gradually disappeared while others coalesced with each other. The structural changes involve the loss of C atoms. These C onions have a larger variety of shapes and sizes as compared with those shown in Fig. 7. The six chosen micrographs capture the dynamics of the fragmentation and reformation of the low density, hollow C onions. The main observations of the electron induced structural changes are: (a) Most C onions are distinct, separated from each other with sizes varying from 20 to100 nm. (b) These onions are strung to the graphitic layers above and below. (c) Amongst a wide range of shapes, the smaller ones appear more spherical than the larger cages. (d) The variations of shapes, density and curvature indicate hollow C onions with non-spherical inner and outer shells. (e) The variable curvature of the corners and edges seem to adjust to the local strains in the closed structures. (f) Majority of the larger ones appear to contain distinctly formed, smaller C onions inside.

*4.3 The coalescing C cages*

The investigations with 120 kV TEM of the C onion structures involving the changes induced by the energetic electrons and the effect of C removal in C onions that are connected with each other are depicted in Fig. 10 (a-b). The structural changes induced in the large (≥100 nm) and the smaller (≤20-40 nm) cages are shown to occur through the coalescence of the smaller cages as can be seen by comparing Fig. 10(a) with 10(b). In Fig. 10(a) two large and many smaller elliptic cages are shown connected with a much larger (~200 nm diameter x 500 nm length) graphitic structure. The smaller cages are emerging out of the larger one. Under intense 120 keV electron beam the smaller onion-like cages on top of the structure in Fig. 10(a) coalesced and produced the larger cages shown in Fig. 10(b). Coalescence of the smaller onions has produced comparatively larger, interconnected, spheroidal structures with variable densities. Another feature of the reformed C onions is that these remain connected with each other. The large kidney shaped cylindrical cage with ~100 nm diameter and 400 nm long can be seen on the left in Fig. 10(a) prior to being irradiated by the electron beam. The irradiation induced changes occurred in its outward convex curvature supplemented by the appearance of two types of nanostructures that are boxed in Fig. 10(b) and enlarged in Fig. 11. The largest structure in Fig. 10 shows diffraction spots from the beam that was focused at centre of the embossed diffraction pattern. One can see that crystallinity of the objects is graphitic with wide range of overlapping planes in the large cage containing many smaller spheroidal, onion-like structures.

*4.4 The emergence of hollow, C cages under 120 keV electron irradiation*

In Fig. 11 the boxed portion of Fig. 10(b) is enlarged and the new emergent structures are shown with arrows. These nanostructures have grown by accumulating the emerging C atoms under electron beam bombardment. Here again the emergence is from within the irradiated graphitic substrate. Out of the six structures identified by arrows, the central ellipsoid pointed by a white arrow is a C onion-inside-onion in the making with ~30 nm x 45 nm dimensions. Black arrows point to five new hollow multiwalled nanotube-like objects with 5-7 shell thickness and different heights ~ 10-40 nm. The 120 keV electrons induced in situ formation of the different stages of the onion and nanotube growth out of the graphitic surface show similarity with similar mechanisms of formation by the C atoms emerging from within the $C_1^+$ irradiated Cu surface.

*4.5 Estimation of the as-grown C onion densities*

Multi-shelled C onions formed as a result of electron irradiation of soot or other high temperature experiments like arc discharge, have been shown to possess structures with $C_{60}$ as the seed around which higher icosahedral fullerene shells are wrapped; the arrangement for a five shelled onion is described as $C_{60}@C_{240}@C_{540}@C_{960}@C_{1500}$ [26]. The essential condition for the adjacent fullerenes is to maintain the inter-shell spacing of 0.334 nm. Such C onions can yield density of graphite (2.26 g cm$^{-3}$). The high energy and intensity electron irradiation of soot [7-10] and the onions produced by energetic C ion implantations in Cu at high temperatures have confirmed the formation of the spheroidal multi-shelled C onions without the hollow inner regions [1-3]. In the present communication, we are reporting the formation of low density C onions formed by the C atoms emerging from the irradiated Cu sheets at room temperature. To estimate such cages' density, let us consider the two sets of the C onions shown in Figs. 7 and 9. The average diameter is about 35 nm. In such an onion one must have about 50 shells with $C_{60}$ at the center with inter-shell separation of 0.334 nm. The total number of C atoms in an onion with n shells is $20n^3$ [Appendix A]. That yields 2.5 x 10$^6$ C atoms in each C onion of 35 nm diameter. Fifty C onions would consume ~10$^7$ C atoms. The total number of the C atoms available in the region where these onions are formed can be worked out by assuming that the irradiated catchment area of the two sets of the C onions (Figs. 7 and 9) is ~1 μm$^2$. On the basis of the total C dose ~10$^6$ μm$^{-2}$, 10$^6$ C atoms would be implanted in 1 μm$^2$ of Cu. A fraction will emerge out; another fraction of the emerging C atoms will be consumed in forming the C onions in addition to the graphitic base underneath. That leads us to the conclusion that at least two orders less C atoms would be available for the C onions that we have observed. This provides us an estimate of the density of C onions ~$10^{-2}$ g cm$^{-3}$. The as-grown C cages formed at room temperature by the accumulation of the emerging C atoms outward from the irradiated Cu are most likely to have low densities resulting in hollow structures.

# 5. Mechanisms for the growth and reformation of the hollow carbon onions

Curvature-related elastic properties of the C shells including the fullerenes, C onions and C nanotubes, can be derived to describe the patterns of growth of the shelled nanostructures. Not only the growth but the reformation of the imperfect spheroids can also be explained under the conditions where C removal from the cages occurs [26]. Spherical shells require the tangential stresses to produce uniformly distributed strain; that only happens in the icosahedral $C_{60}$ grown in high temperature environments or in the well annealed, multi-shelled carbon onions [1-7]. Deviations from perfect sphericity as would occur in the case of $C_{240}$, $C_{540}$ and the larger free standing fullerenes, have been explained in the nanoelastic C cages by including stretching of the cage shells as the first order and bending as the second order effects. The departure from perfect sphericity is illustrated graphically in Fig. 12(a). It can be considered a typical large (>$C_{60}$) icosahedral fullerene with (twelve) outward protruding corannulenes superimposed on an inner shell of radius $R$ with shell thickness $t$. The magnitude of the protrusion $\zeta$ is the difference between the circumscribing and inscribing radii of the appropriate Goldberg polyhedral spheres. Fig. 12(a) is the graphical representation of an otherwise perfect sphere that has built-in protrusions. The role played by these outward protrusions was discussed elsewhere in the context of the growth habits of the carbon nanotubes [27], here we present the essential features from the model that are relevant for describing the growth and reformation mechanisms of C nanostructures with emphasis on the onion-like cages shown in Fig. 11. The departure from perfect symmetry in the form of the protruding curved surfaces emerging out of the spherical surfaces can be defined by two additional parameters as indicated in Fig. 12(a); the width of the meridianal strip $d\sim\sqrt{tR}$ that surrounds the outward bulge and the local radius of curvature ($1/r$). The emergence of the protruding bulge indicates the existence of an in-built internal force whose magnitude can be determined from the extent of the bulge (Fig. 12(b)). This outward lifting force $f_o$ can be evaluated by considering the bending and stretching energies in the defect volume $\zeta^2 R$. It is evaluated as $f_o \approx Yt^{5/2}(\zeta^{1/2}/R)$, where $Y$ is the elastic modulus. We can derive an expression for the measure of the protrusion $\zeta$ to relates the internal stress $P$ with the curvature in the defect volume and obtain $\zeta \approx Y^2 t^5/(R^4 P^2)$. In this equation $\zeta$ increases when $P$ reduces leading to an inverse relationship between $\zeta$ and $P$. Thus an unstable equilibrium is established where the bulges with large $\zeta$ grow on their own while the smaller ones shrink. This happens at the stages of C accretion and removal from the cages. A critical value of $P$ exists for $\zeta \approx t$ beyond which even small changes in the shape of the shells increase spontaneously; $P_{cr} \approx Yt^{5/2}/(\zeta^{1/2} R^2)$ [27]. In Fig. 12(b) the effect of $f_o$ on an otherwise perfect spherical shell is shown and its quantitative value for the fullerenes with large radii and the associated pentagonal protrusions superimposed on an internal sphere with radii $R$ is shown in Fig. 12(c). Also plotted in 12(c) is the critical stress $P_{cr}$ as a function of the large fullerenes' radii. The slowly varying outward lifting force $f_o$ has $R^{-1}$ dependence. This force is effective in conditions where the initial curved surfaces start to form and are lifted outwards. In the case of cages with protrusions the critical stress $P_{cr}$ reduces as $R^{-2}$. The outward protruding bulges associated with large spherical structures lead to the buckling of the cages at smaller critical stresses. Hence the stable growth of the C cages, nanotube-like or onion-like, depends on the cage thickness ($t$) and the local radius of curvature ($1/r$) that eventually defines $\zeta$. The above mentioned model was developed for large, independent fullerenes (>$C_{60}$) but it is valid for all the cages with volume defects like the outward bulging surface protrusions. The appropriate thickness $t$ and the value of Young's modulus $Y$ will have to be invoked to get realistic estimates of the forces and the critical stresses. C removal is likely to transform the protrusions or the bulges of the cage with smaller radii into larger ones. The net effect of the changes induced is to stabilize the structures. Prolonged electron irradiations tend to remove the excessive strains introduced by the variable radii within the same cage –as was seen in the structural changes of Fig. 10(a)→10(b).

The C cages shown in Figs. 3-10 can be understood by comparing with the energetic electron induced transformations. In Fig. 13(a) the in situ grown structures shown in Fig. 11 are reproduced with the background removed. Here two distinct but structurally related C nanostructures are formed by the C atoms that are released by 120 keV electrons from the underlying graphitic surface. shown in 13(a). The graphical representation of the growth mechanisms of such nanostructures is presented in four sequences 1-4 in Fig. 13(b). The structure underneath the irradiated region and the rate of flow of the emerging C atoms determine the curvature and the lateral extent of the newly forming C layers (step 1). For a steady flow of C, shells-inside-shells grow without significant variations in diameter resulting in the multishelled nanotube-like features (step 2). For increasing outward flow of C one is likely to obtain onion-like structure (step 3). Discontinuous or variable C flow rates are likely to yield onion-inside-onions (step 4). Majority of the C onions seen in Figure 6 belong to this category. Such an object can also be seen in the TEM micrograph Fig. 10(a). In situ growth of an onion-inside-onion and the six hollow nanotubes is proposed to be due to the C atoms released by the 120 keV electrons from the outer surface of the large graphitic cage. One finds similarities by comparing the C onions formed by C atoms which emerge from the irradiated Cu surfaces or the ones that are released by breaking the $sp^2$ bonds of the graphitic structures by 120 keV electrons. We believe that the latter explains the routes to the cage formation of the former.

## 6. Density comparison of onions

We have introduced a new technique of $sp^2$-bonded nanostructure formation where the C atoms emerging from within the $C_1^+$ irradiated Cu produce a wide range of graphitic configurations. Low density, multiple layered structures and a broad range of C cages with sizes from few to tens of nm have been observed. The experimental evidence was provided for a wide variety of C nano cages with different shapes, sizes and shell thickness by accumulating the emerging C atoms out of the irradiated Cu grids and sheets. The variety of C cages include the partially open cages, the closed cage structures, C onion-inside onions and those cages that were in various stages of their growth. Larger caged structures with dimensions ~few hundred nm produced by the coalescence of smaller cages were shown. The structure formation occurs at room temperature in high vacuum without the presence of any catalyst (e.g. Fe, Co or Ni). Formation of the C cages at room temperatures forbids the Stone Wales transformations by C–C bond rearrangements to occur [28]. This rearrangement can smooth out the sharp corners resulting from the abutting pentagons in fullerenes and the hemispherical ends of the C nanotubes. It can occur with thermal activation [1-3, 9-16], or without thermal activation by the energetic electron beam-induced breaking and remaking of the C–C bonds [4-7, 22]. However, the density of the C cages achieved by annealing or electron beam irradiation of soot is of the order of graphite density (~2 g cm$^{-3}$). Our technique, on the other hand, produces C cages with three orders less dense; therefore, the C removal under 120 keV electrons in our TEM did not produce the C onions with shell–inside–shell of the type $C_{60}@C_{240}@C_{540}@C_{960}@C_{1500}@$.... Under electron irradiation the C onions (Figs. 9-11) changed their shapes and reduction in the sharpness of the corners occurred. Cages coalesced but the high density; perfect spherical C onions were not produced. This is due to the inherent low density of the cages and the absence of thermal activation. The rate of the $C_1^+$ irradiation and the routes that C atoms take to escape from within the Cu lattice determine the nature and type of the C onions and the graphitic layers that are formed. The spherical and cylindrical curvatures are seen as the essential features of the nanoelastic response of the cages by which the C onions and the graphitic layers accommodate variations in the C accretion rate and the Cu surface irregularities in and around the region of C emergence. Patterns of the cage formation yield the diversity of the curvature in the C onions that have been presented in this communication. We have proposed a growth mechanism for the hollow, irregular shaped C cages by utilizing a model developed to describe the growth habits of the large fullerenes and C nanotubes [26,27]. The in situ growth of C onion-like and C nanotube-like structures with 120 keV electron irradiation of a large graphitic structure has similarities with the cages grown by the accretion of C emerging from the irradiated Cu. The experimental evidence and the model based on the elastic properties of the nano C cages were used to describe the growth and the reformation of the C onions.

## 7. Conclusion

Evidence for the formation of low density, irregular shaped, hollow C onions on $C_1^+$ irradiated Cu grids and sheets at room temperature was presented. The routes and patterns of formation and transformation of these cages were investigated with Confocal, scanning and transmission electron microcopies. Curvature of the closed cages seems to be the natural outcome of the growth of the atom-by-atom accretion of C. The $C_1^+$ irradiation rates determine the shape, size and the number of shells of these onion-like structures. $C_1^+$ irradiation doses ~ $10^{15}$ cm$^{-2}$ with irradiation rates ~$10^3$ cm$^{-2}$ s$^{-1}$ yield widely dispersed graphitic cage growth along the edges of the irradiated Cu. The outcome of variable C accretion rates seems to produce large ensembles of multishelled closed cages with variable radii and shapes. The individual C onions are shown to gain or lose C atoms under energetic electron irradiation with the changes of shape, density and radius. Smaller cages were seen to coalesce to form larger ones. In situ formation and transformation of cages under 120 keV electrons identify the mechanisms of growth and cage shrinkage. Possible applications of such low and maneuverable density C onions may be to produce nanostructures with variable density of states (DOS) having selective absorption and emission properties. The hollow interiors could be filled with suitable filling materials to be used as quantum dots. The effect of high temperature annealing on selective cages can be used to study formation of cages with specific characteristics like outer and inner cage diameters, shell thicknesses and adhesion to graphitic sheets, for example.

# References

[1] Cabioc'h T, Riviere JP, Delafond J. A new technique for fullerene onion formation. J Mater Sci 1995;**30**(19):4787-92.

[2] Cabioc'h T, Thune E, Rivie're JP. Structure and properties of carbon onion layers deposited onto various substrates. J Appl Phys 2002;**91**(3):1560-7.

[3] Abe H. Nucleation of carbon onions and nanocapsules under ion implantation at high temperature. Diam Relat Mater 2001;**10**(8):1201-4.

[4] Iijima S. Direct observation of the tetrahedral bonding in graphitized carbon black by high-resolution electron microscopy. J Cryst Growth 1980;**50**(3):675-83.

[5] Ugarte D. Curling and closure of graphitic networks under elec-tron-beam irradiation. Nature 1992;**359**(6397):707-9.

[6] Ugarte D. High temperature behavior of "fullerene black". Carbon 1994;**32**(7):1245-48.

[7] Ugarte D. Onion like graphitic particles. Carbon 1995;**33**(7):989-93.

[8] Xu B , Tanka SI. Formation of giant onion-like fullerenes under Al nanoparticles by electron irradiation. Acta Mater 1998;**46**(15):5249-57.

[9] Saito Y, Yoshikawa T, Inagaki M. Growth and structure of graphitic tubules and polyhedral particles in arc-discharge. Chem Phys Lett 1993;**204**(3-4):277-82.

[10] Sano N, Wang H, Chhowalla M. Synthesis of carbon onions in water. Nature 2001;**414**(6863):506-7.

[11] Sano N, Wang H, Alexandrou I. Properties of carbon onions produced by an arc discharge in water. J Appl Phys 2002;**92**(5):2783-88.

[12] Guo JJ, Wang XM, Yao YL. Structure of nanocarbons prepared by arc discharge in water. Mater Chem Phys 2007;**105**(2-3):175-8.

[13] Tatiana G, Sabine U, Fritz F. Carbon onions produced by laser irradiation of amorphous silicon carbide Chem Phys Lett 2003;**373**(5-6):642-5.

[14] Radhakrishnan G, Adams PM, Bernstein LS. Plasma characterization and room temperature growth of carbon nanotubes and nano-onions by excimer laser ablation. Appl Surf Sci 2007;**253**(19):7651–5.

[15] Grieco W J, Howard J B, Rainey L C. Fullerenic carbon in combustion-generated soot. Carbon, 2000, **38**(4): 597-614.

[16] Johnson MP, Donnet JB, Wang TK. A dynamic continuum of nanostructured carbons in the combustion furnace. Carbon 2002;**40**(2):189-94.

[17] Li BY, Wei BQ, Liang J. Transformation of carbon nanotubes to nanoparticles by ball milling process. Carbon 1999;**37**(3):493-7.

[18] Huang JY, Yasuda H, Mori H. Highly curved nanostructures produced by ball milling. Chem Phys Lett 1999;**303**(1-2):130-4.

[19] Oku T, Narita I, Nishiwaki A. Formation, atomic structural optimization and electronic structures of tetrahedral carbon onion. Diam Relat Mater 2004;**13**(4-8):1337-41.

[20] Szerencsi M, Radnoczi G. The mechanism of growth and decay of carbon nano-onions formed by ordering of amorphous particles. Vacuum 2010;**84**(1):197-201.

[21]    Blank VD, Kulnitskiy BA, Perezhogin I. A. Structural peculiarities of carbon onions, formed by four different methods: onions and diamonds, alternative products of graphite high pressure treatment. Scripta Materilia 2009; **60**(6):407-10.

[22]    Banhart F. Irradiation effects in carbon nanostructures. Rep Prog Phys 1999;**62**(8):1181- 221.

[23]    Mattevi C, Kim H, Chhowalla M. A review of chemical vapor deposition of graphene on copper. J Mater Chem 2011;**21**(10):3324-34.

[24]    Fuks D, Van Humbeeck J, Liubich V. Carbon in copper and silver: diffusion and mechanical properties. J Mol Structure (Theochem) 2001;**539**(1-3):199-214.

[25]    Sun J, Zhang MD. Interface characteristics and mechanical properties of carbon fibre reinforced copper composites. J Mater Sci 1991;**26**(21):5762-6.

[26]    Ahmad S. Continuum elastic model of fullerenes and the shpericity of the carbon onion shells. J Chem Phys 2002;**116**(9):3396-3400.

[27]    Ahmad S. Criteria for the growth of fullerenes and single-walled carbon nanotubes in sooting environments. Nanotechnology 2005;**16**(9):1739-45.

[28]    Stone AJ, Wales DJ. Theoretical- studies of icosahedral $C_{60}$ and some related species. Chem Phys Lett 1986;**128**(5-6):501-3.

**Acknowledgements**

Authors acknowledge financial support by the Higher Education Commission of Pakistan (HEC) to set up the 2 MV Pelletron and electron microscopy labs at Government College University (GCU), Lahore. We acknowledge Mr. M. Khaleel for technical support with the Pelletron and Ms. Farwa Nurjis of NIBGE, Faisalabad for the Confocal Microscopic images.

**Appendix A**

The calculation of the total number of C atoms in an onion can be done by adding the C atoms in each shell. The number of C atoms in the icosahedral fullerene shell $C_{60}$, $C_{240}$, $C_{540}$,.…. follow the empirical formula $N_n=60n^2$, where n identifies the fullerene shell; n=1 for $C_{60}$, 2 for $C_{240}$, 3 for $C_{540}$ and so on. For a C onion of the type $C_{60}@C_{240}@C_{540}@C_{960}@C_{1500}@C_{2160}$...The sum of the atoms in all shells of the onions is $\Sigma N_n=60\Sigma n^2=60(1/6(n+1)(2n+1)))\approx 20n^3$ (for n>10). Such an onion has the density of graphite ~ 2 g cm$^{-3}$.

**Figure Captions**

Fig. 1- SEM micrographs of the irradiated Cu grid and a sheet. (a) 10 μm thick Cu grid with 90 μm diameter holes, irradiated with 0.2 MeV $C_1^+$ show the circular edges decorated with C microstructures. Inset: Confocal microscopic image of the grid shows bright C objects along the circular edges. (b) A 600 μm long object emerging out of the inner edge of a 3 mm diameter hole in 120 μm thick irradiated Cu sheet. Inset: The same object in its Confocal image against the dark Cu background.

Fig. 2- The Confocal microscopic images of the irradiated Cu grids and sheets with edges in (a) and (b) show areas ~ 1mm$^2$ from the $C_1^+$ irradiated Cu grids with square and circular holes. (c) The SEM micrograph showing the irradiated 120 μm thick Cu sheet with 3 mm diameter hole; the circular edge are decorated with the growth of C microstructures. Inset: A selected portion with the Confocal image showing bright C objects grown along the edge.

Fig. 3- The scanning and transmission electron microscopic images of the C micro and nanostructures. (a) A SEM micrograph of the irradiated 120 μm thick Cu sheet with 1 mm diameter hole showing the highly decorated edge with large objects emerging out of the edge. (b) TEM image of the circled portion in (a) showing a 10 μm x 15 μm graphitic structure. The boxed regions numbered 1 and 2 are explored in the later figures.

Fig. 4- The two selected regions of the graphitic object shown in Fig. 3(b) are shown as TEM micrographs in 4(a) and 4(c). These are the images of the two outward protruding graphitic structures with their diffraction pattern shown in (b) and (d), respectively.

Fig. 5- The C onions of a variety of shapes strung on graphitic layers. (a) An area of ~ 1μm$^2$ shows a rich variety of inter-connected graphitic layers studded with large and small cages. Two boxed regions are pointed for enlargement. (b) The white boxed region is enlarged to show an assembly of a large number of C onions. (c) A large collection of ~100 C onions is connected to a graphitic network of layers. (d) The diffraction pattern from the centre of the onions sheet in 5(c) showing large number of spots indicative of the graphitic crystallinity.

Fig. 6- A stack of the inter-connected graphitic layers. (a) A large graphitic stack of volume~5 x 10$^7$ nm$^3$ shows two hollow, hemispherical nano shells protruding outwards. These are further enlarged in (b). The outer shells of ~3 nm thickness highlight the inner hollow regions.

Fig. 7- A batch of the inter-connected C onions emerging from a graphitic base; the average diameter is about 30±10 nm. (a) The upper halves seem in higher state of spherical perfection and are closed. (b) The partially open lower halves of the C onions are shown. Inset: An enlarged view that highlights the structural features of these C onions.

Fig. 8- The effect of the continuous, intense 120 keV electron beam irradiation on the batch of outward protruding onions shown in Fig. 7 is displayed here 8(a)–(f). The changes in the assembly as a whole and in the individual C onions are due to the knocking off of C atoms and breaking of the intra-onion bonds.

Fig. 9- The electron beam irradiation damage to a large number of hollow C onions spread perpendicular to the beam direction shown in six consecutive TEM micrographs 9(a)-(f). These micrographs were obtained at regular intervals during the 1200 s after the first one. Large displacements, modifications, coalescence and the annihilations can be identified on close examination. The 265 nm diameter white circle in the micrographs is to guide the eye and to highlight the ongoing structural changes.

Fig. 10- The electron induced coalescence of the smaller C onions into larger ones and the shape changes of the larger graphitic cages. (a) A collection of three large (≥50 nm) and many smaller C onions, all sticking together in their pristine positions is shown in the first TEM micrograph. (b) After intense 120 keV electron beam irradiation in TEM for 600 s the smaller ones coalesced into larger C onions that have relatively well defined surfaces. Inset: the white box has the diffraction spots from the center of the large graphitic cage.

Fig. 11- In situ formation of the hollow onion-like and nanotube-like structures. The boxed region in Fig. 10(b) is further enlarged to reveal the six new, multi-walled C nanotubes of ~2-3 nm thick shells. These emerged out of the reshaped, larger graphitic caged structure under 120 keV electron irradiations. Inset: shows the diffraction spots from the centre of the large cage.

Fig. 12- The model of a nanoelastic icosahedral cage is shown in (a) with outward protruding regions along the axes of the corannulenes of a typical large fullerene superimposed on a central spherical cage of radius $R$; $\zeta$ –the measure of the protrusion, $t$ –the shell thickness, $r$ –the local protrusion radius, $d$ –thickness of the strip around the protruding region containing the major part of the elastic energy. (b) shows the outward lifting force $f_0$ that is plotted along with the critical stress $P_{cr}$ in (c) for a range of large fullerene shells.

Fig. 13- (a) The in situ growth of the nanotube-like and the onion-like structures in Fig. 11 is reproduced with the background removed to relate the new growth with the graphical model presented in (b). (b) Four proposed stages (1-4) of formation of the multi-walled C nanotubes or multi-shelled C onion-inside-onion are graphically represented. The emergence of C from under the Cu surface is shown to define the mechanism of formation.

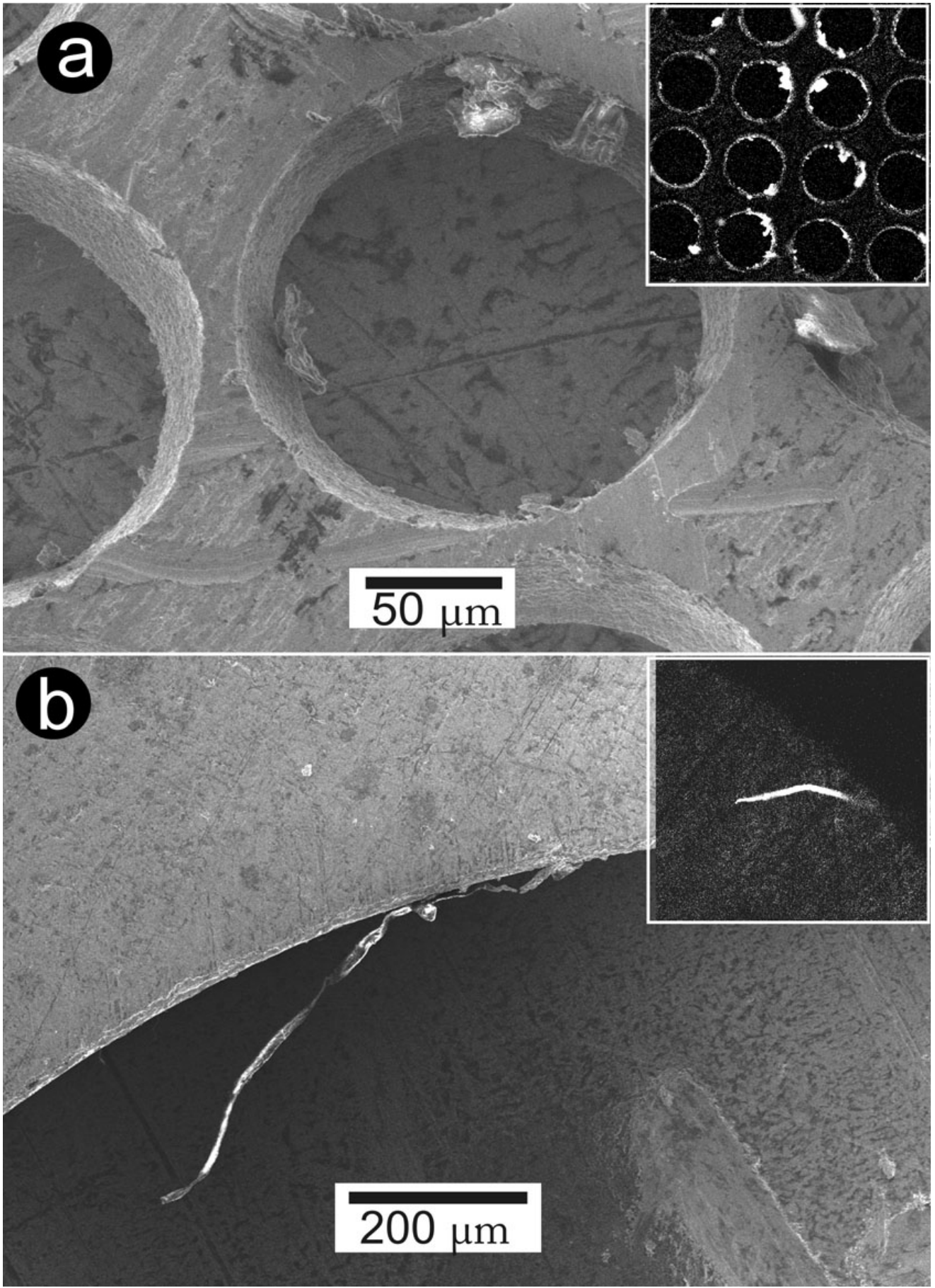

**Fig. 1.**

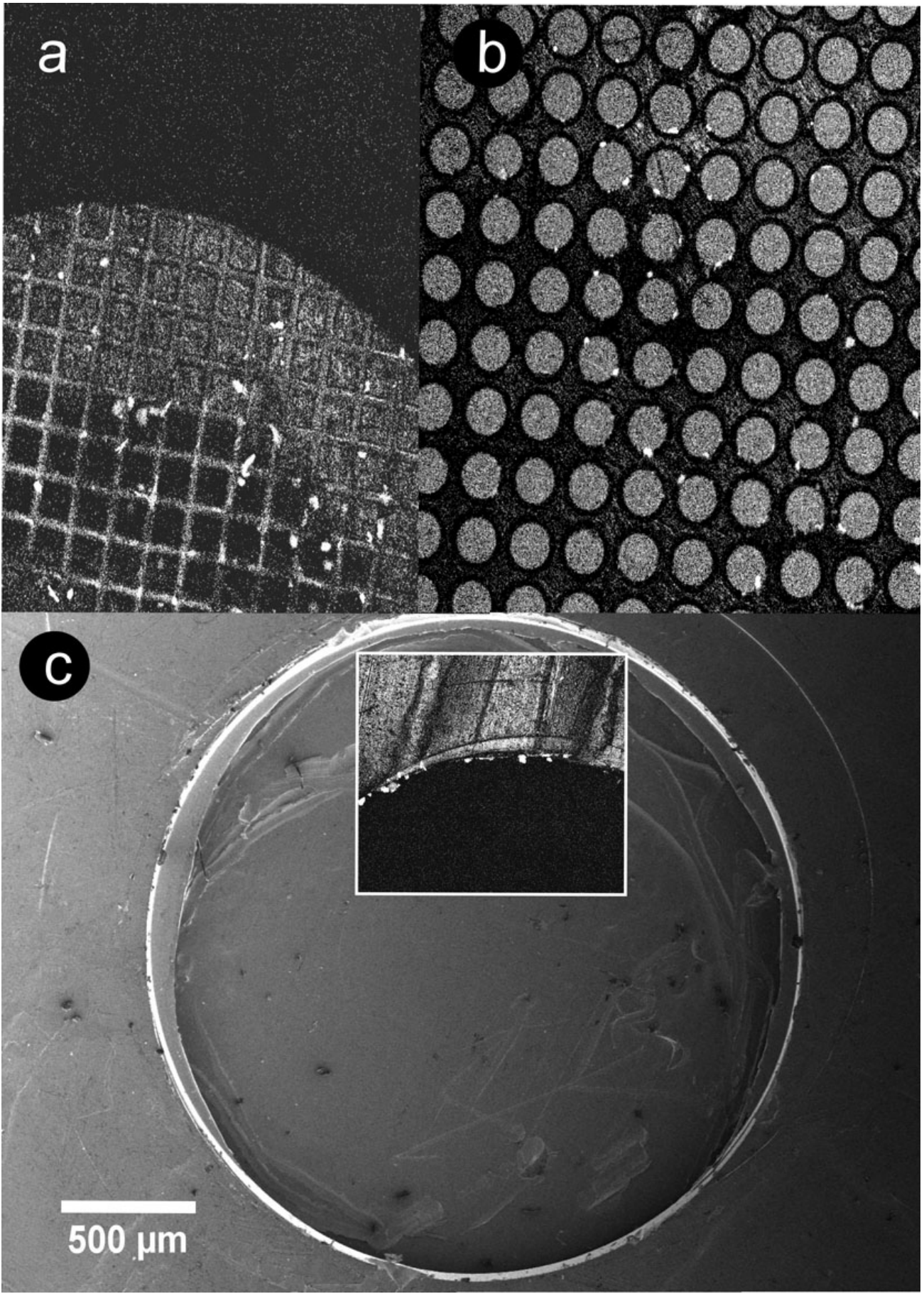

**Fig. 2**

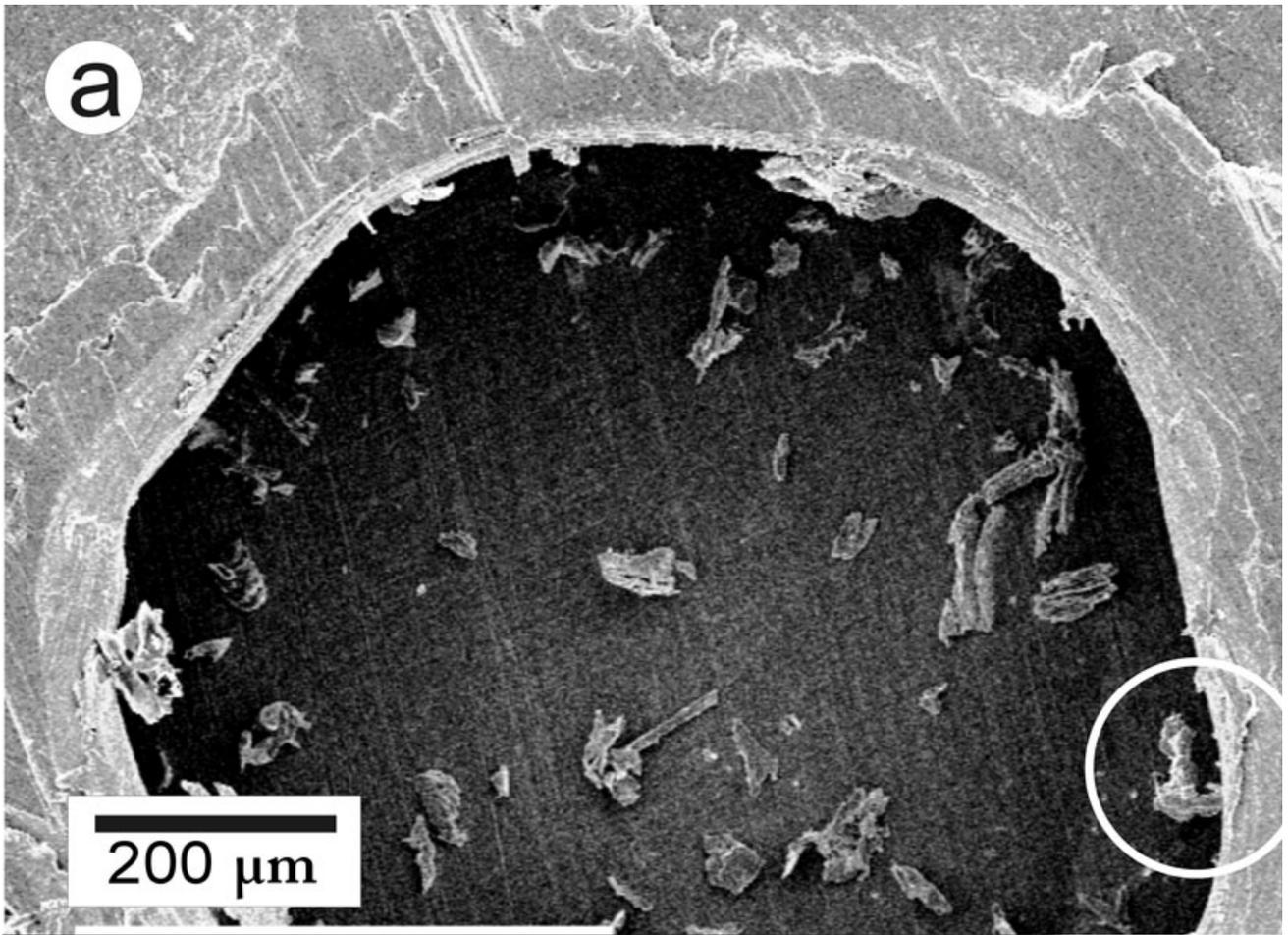
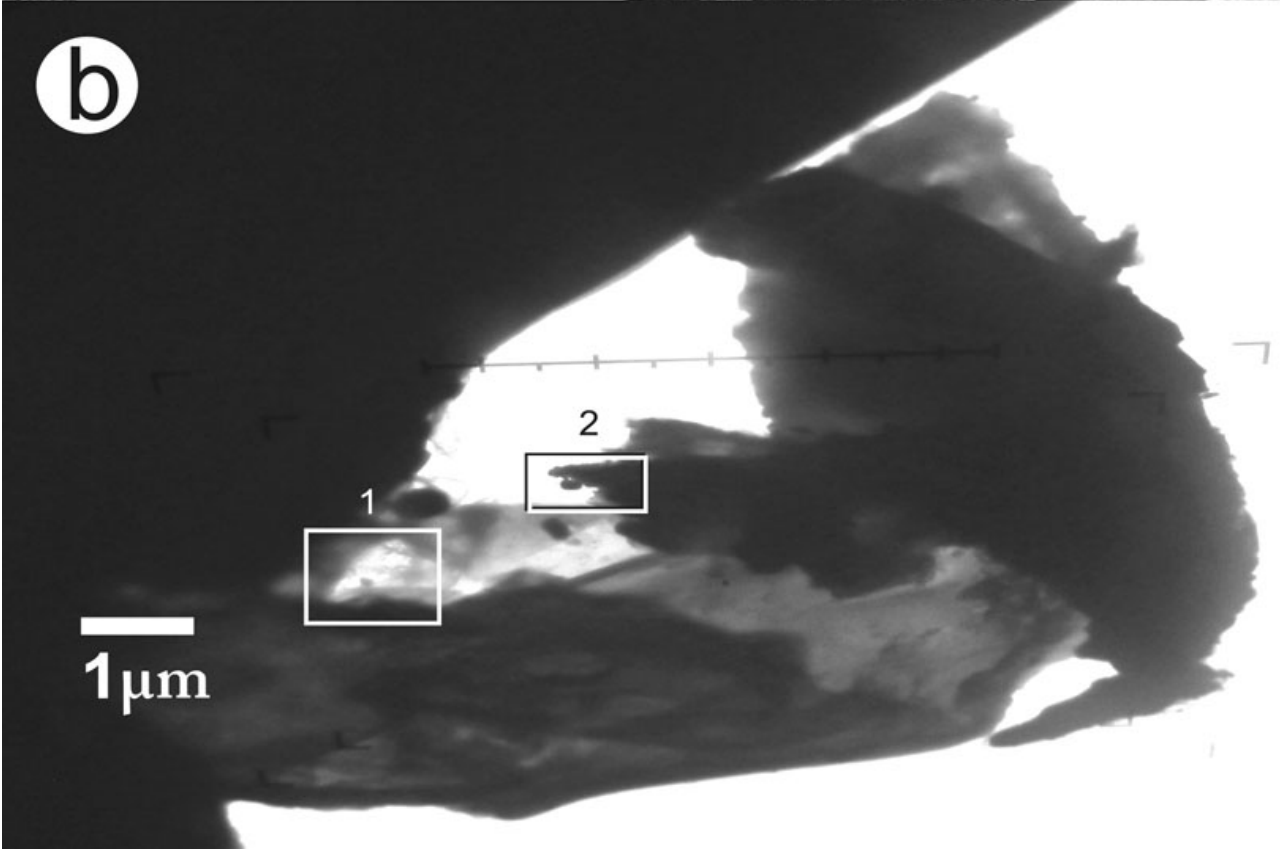

**Fig. 3**

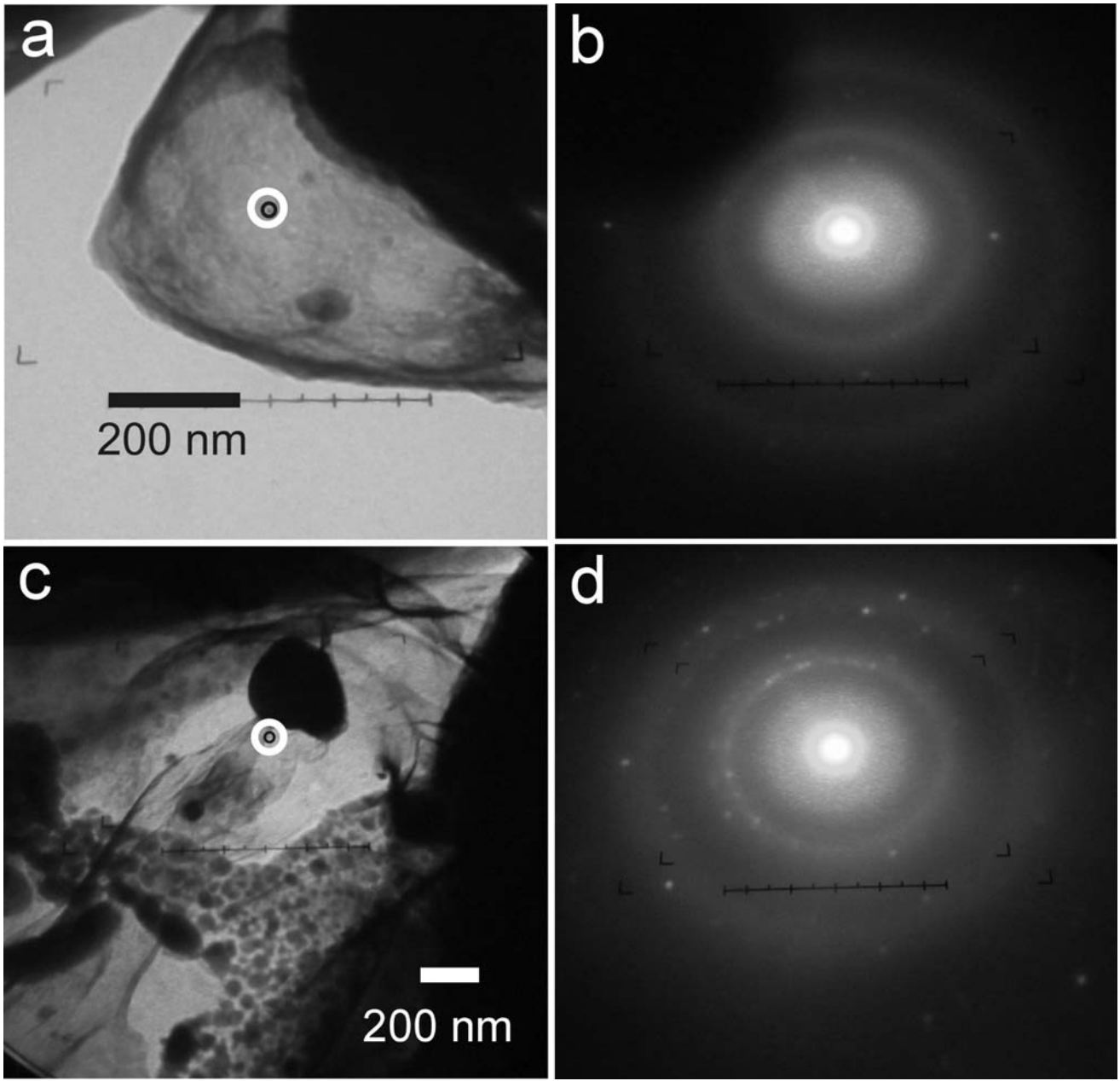

**Fig. 4**

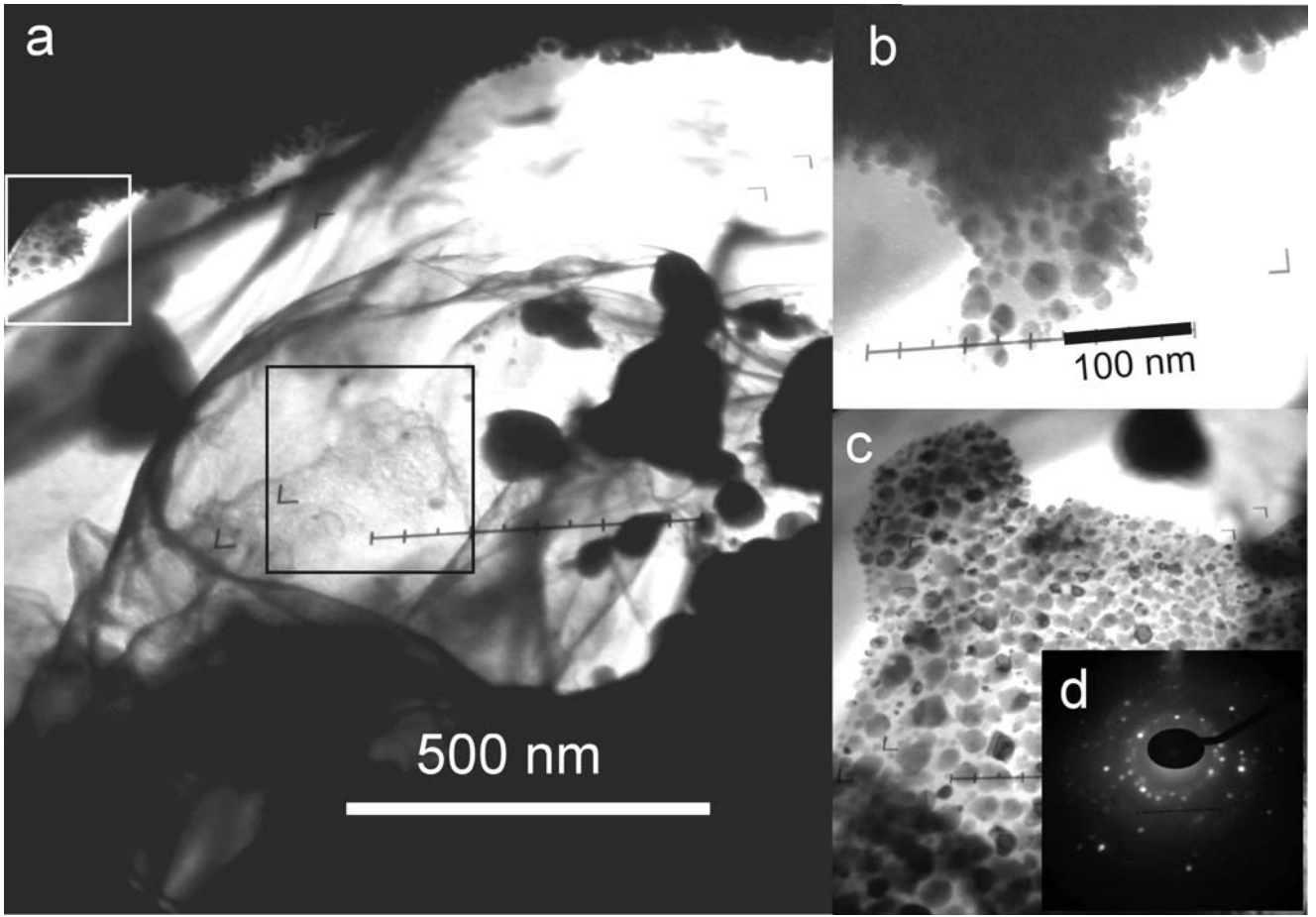

**Fig. 5**

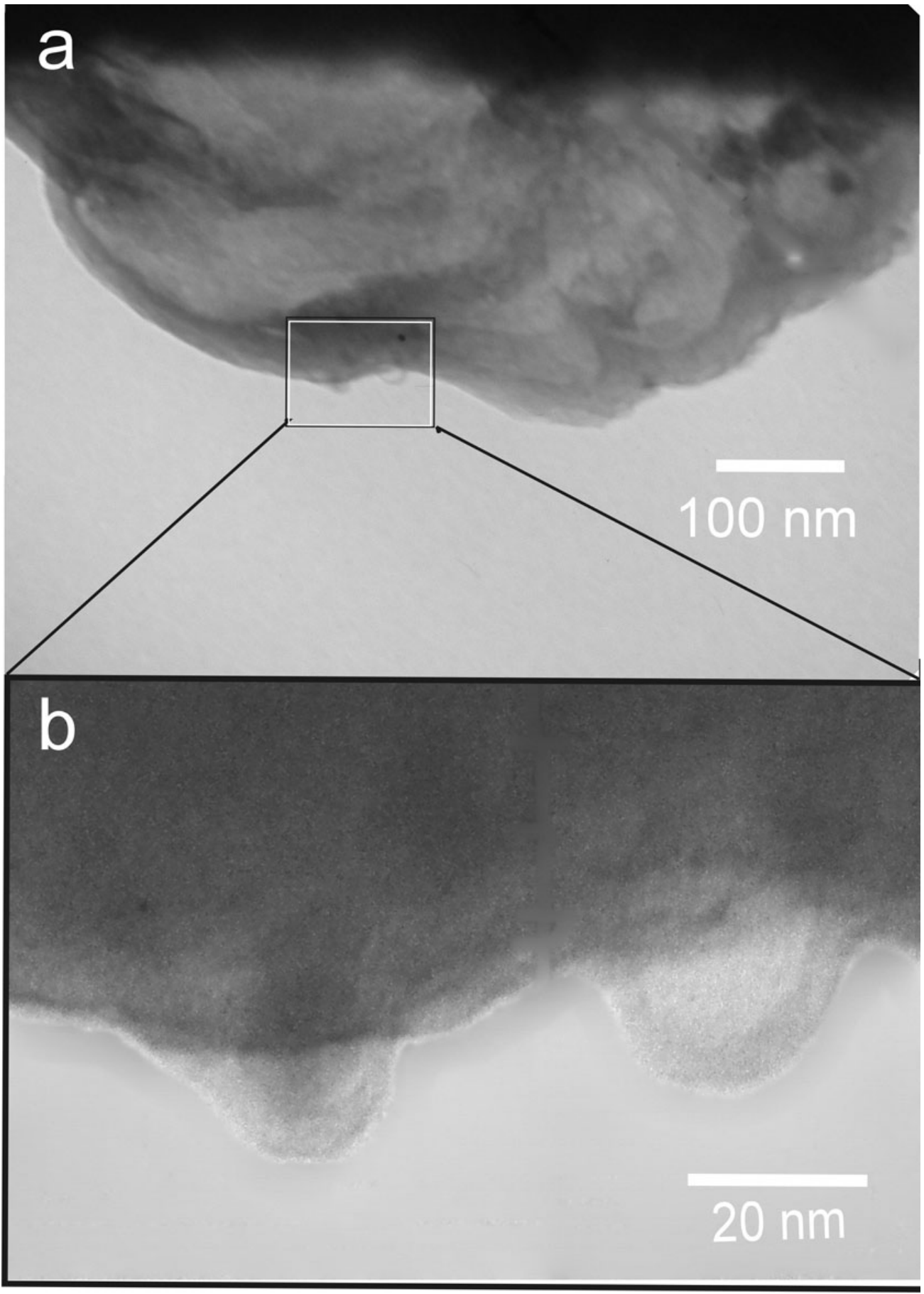

**Fig. 6**

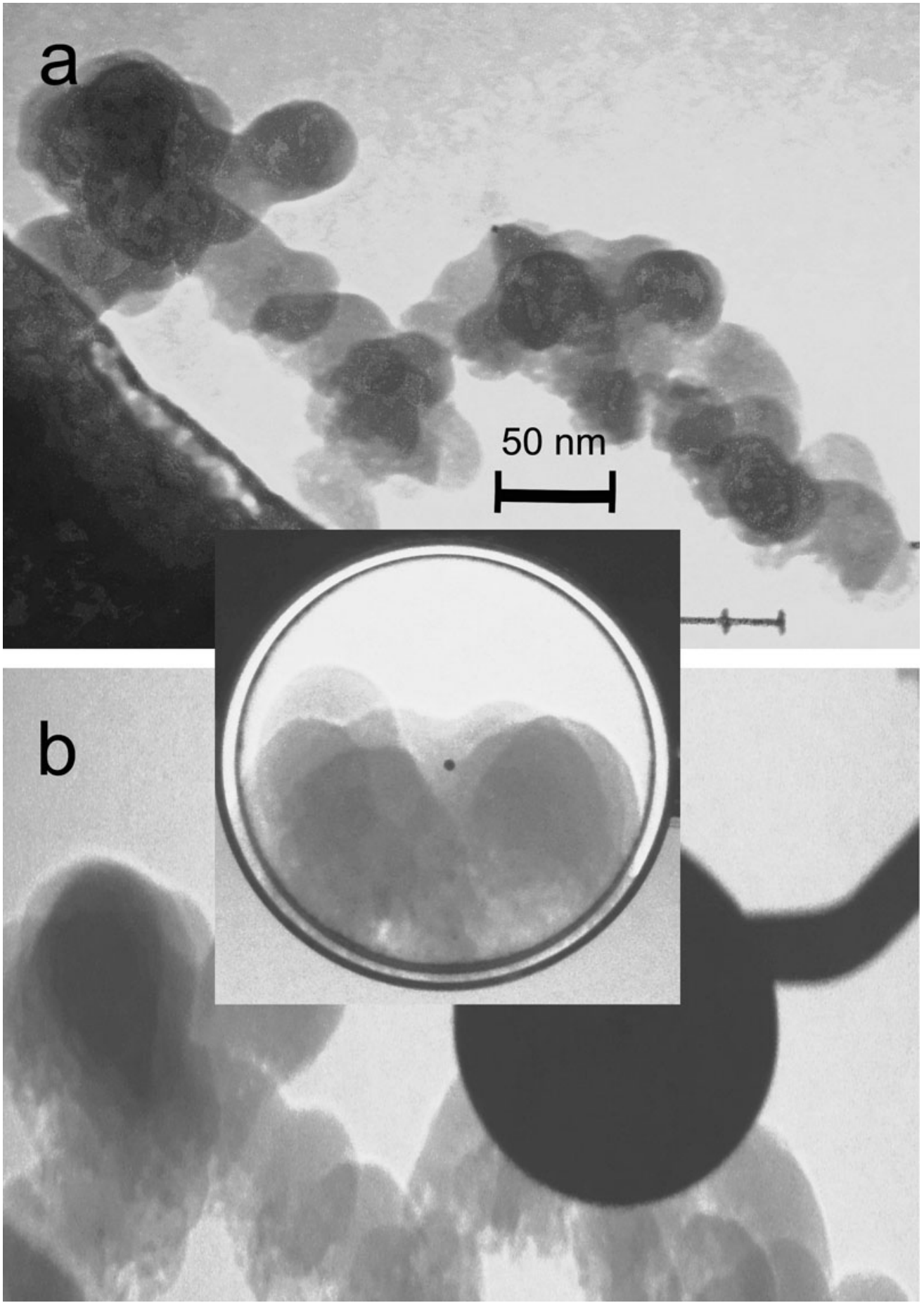

**Fig. 7**

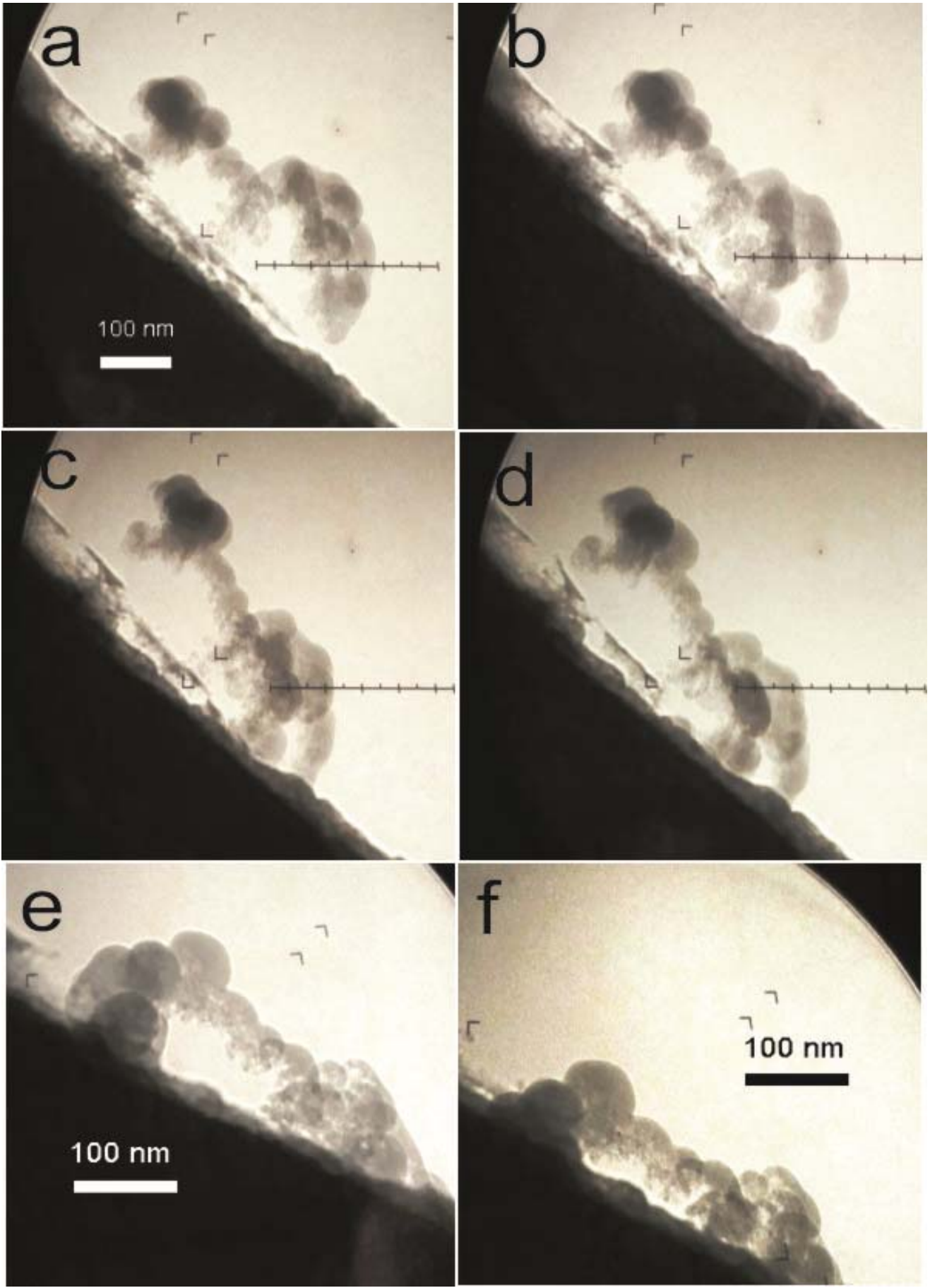

**Fig. 8**

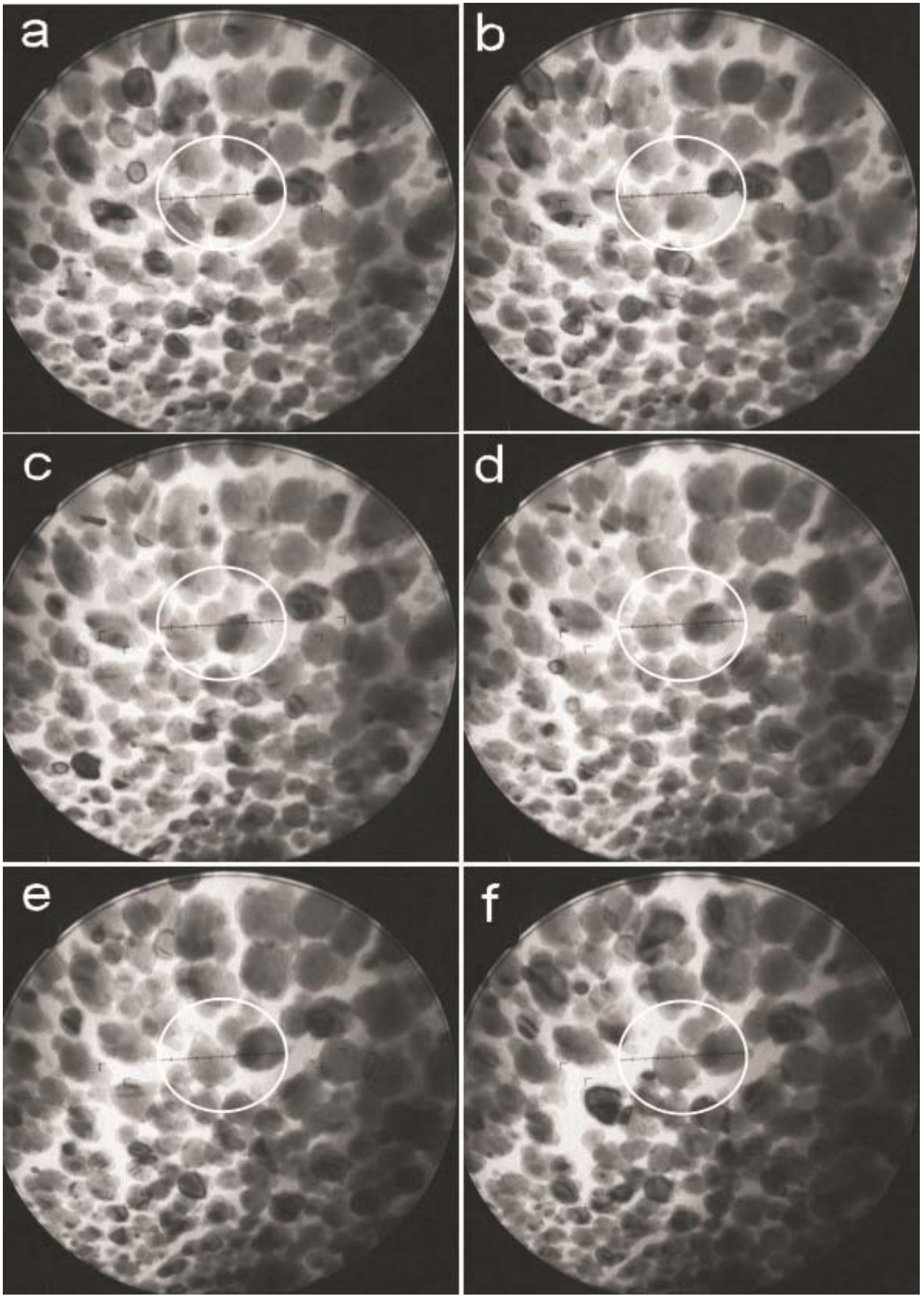

**Fig. 9**

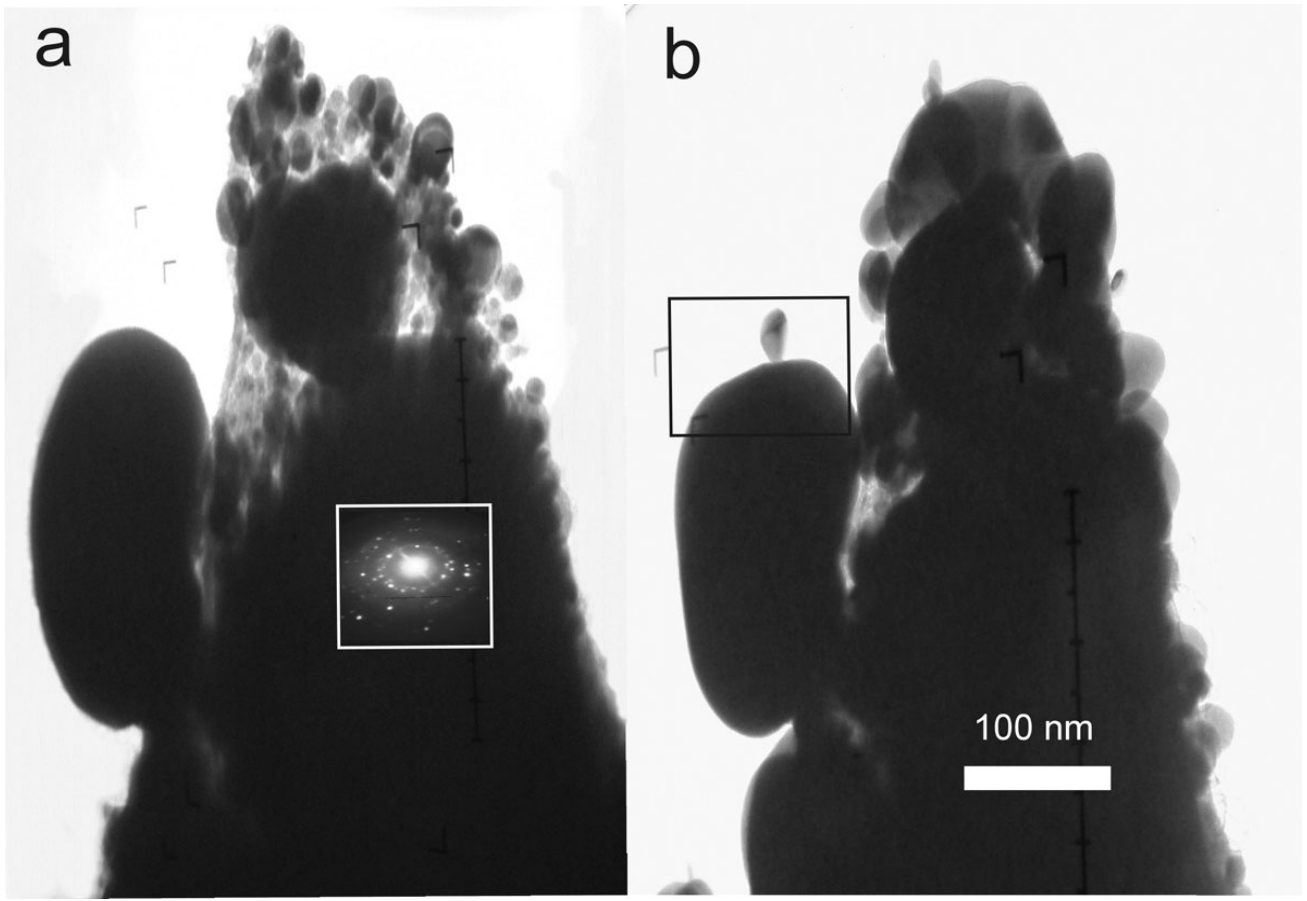

**Fig. 10**

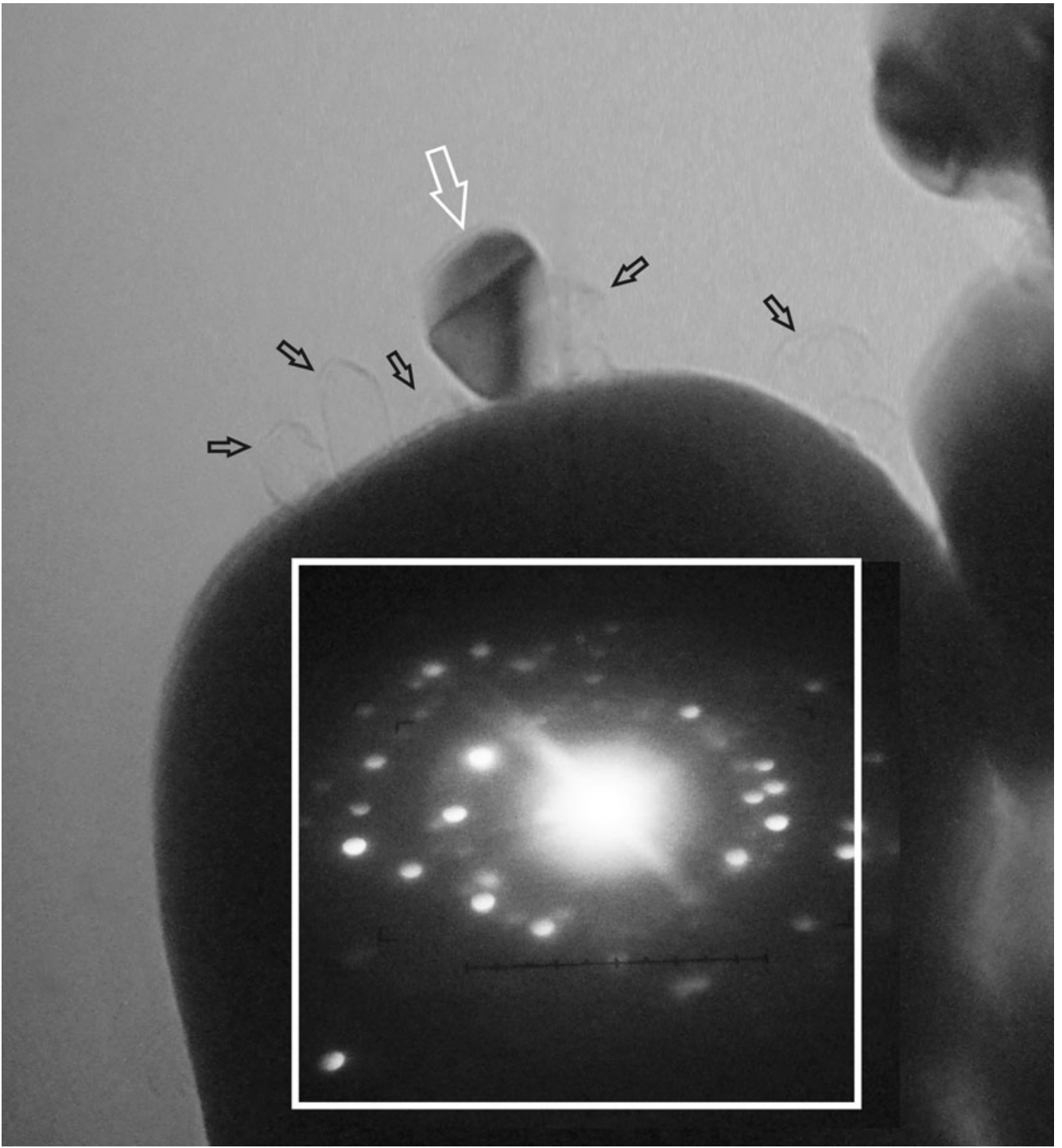

**Fig. 11**

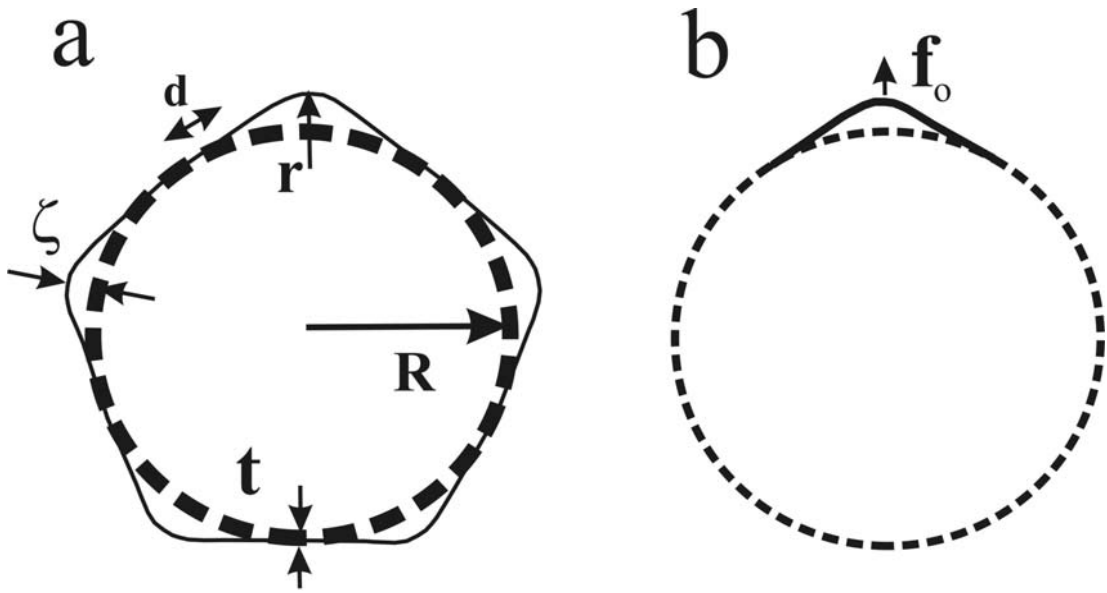

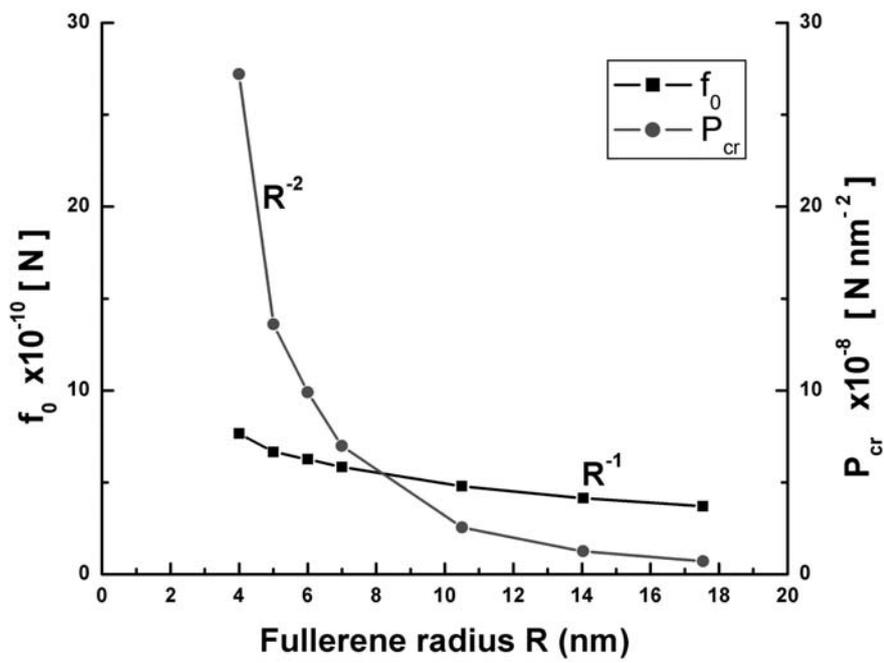

**Fig. 12**

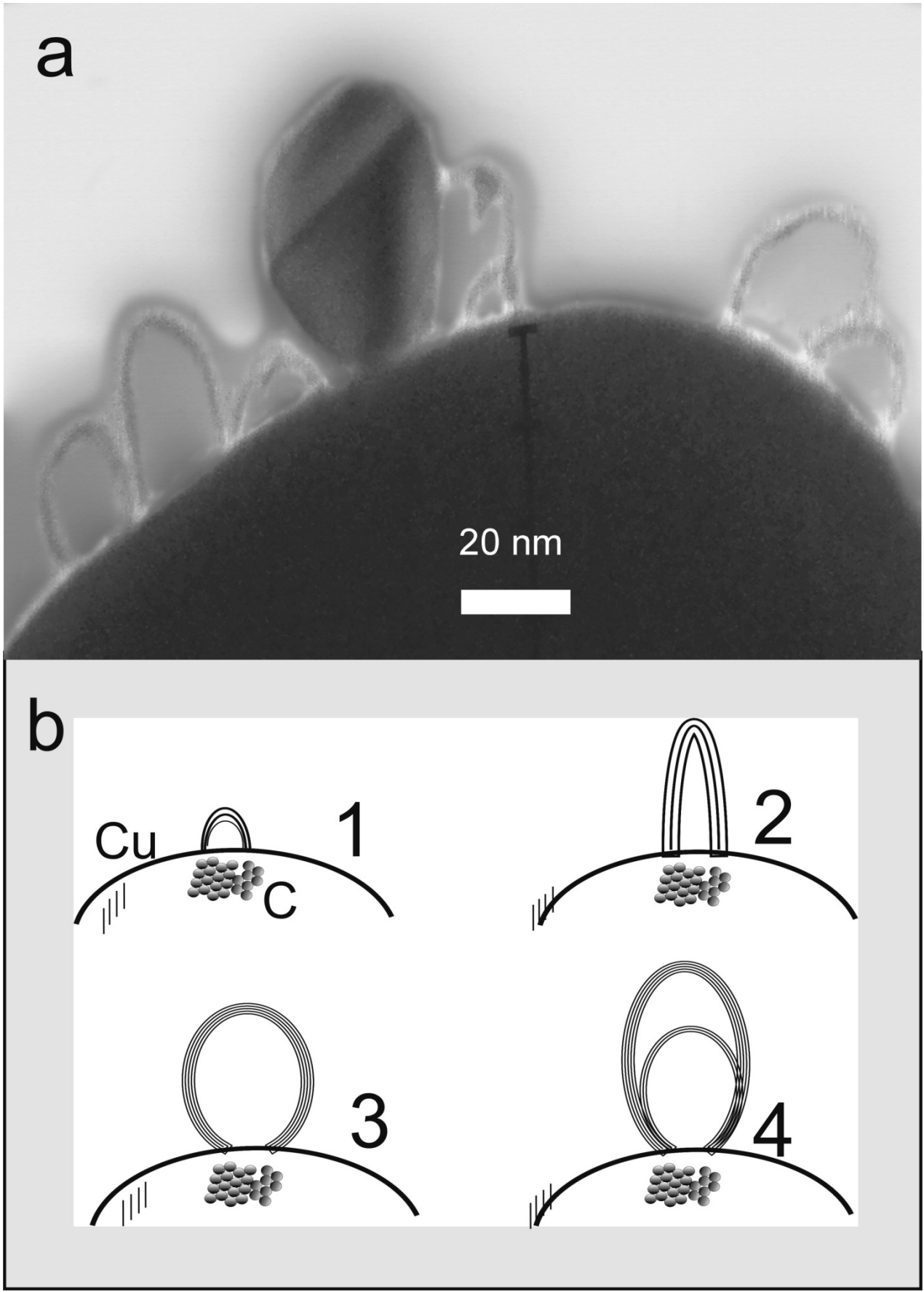

**Fig. 13**